# ELABORATE LEXICON EXTENDED LANGUAGE WITH A LOT OF CONCEPTUAL INFORMATION

Jean Luc Razafindramintsa[1], Thomas Mahatody[2] and Josvah Paul Razafimandimby[3]

[1]Research Institute School for Computer Modelisation, Laboratory for Mathematical and Computer Applied to the Development Systems, University of Fianarantsoa Madagascar
[2]National School for Computer Engineering, University of Fianarantsoa Madagascar
[3]Laboratory for Mathematical and Computer Applied to the Development Systems, University of Fianarantsoa Madagascar

## ABSTRACT

*The use of model such as LEL (Lexicon Extended Language) in natural language is very interesting in Requirements Engineering. But LEL, even if it is derived from the Universe of Discourse (UofD) does not provide further details on the concepts it describes. However, we believe that the elements inherent in the conceptual level of a system are already defined in the Universe of Discourse. Therefore, in this work we propose a more elaborate natural language model called eLEL. It is a model that describes the concepts in a domain in more detail than the conventional LEL. We also propose a modeling process of a domain using an eLEL model.*

## 1. INTRODUCTION

Elucidation, modeling and analysis are both the inherent and the basic elements of requirements engineering [9]. In order to obtain the essential information and to understand the elucidation phase problems, [15] stipulate the use of several resources by requirements engineers including the analysis and careful reading of sources of information documents. For this analysis and reading, source of information documents such as companies businesses, surveys, interviews, the regulation or related text and other sources of information systems [15] should be assessed. The documents generated during this phase of elucidation (investigation reports, interviews, regulation or related text and other sources of information systems) are written in natural language. These documents contain the terms in the domain of the problem which are the terms used by customers and users [11, 13]. These documents are then used by relevant professionals, considering that professionals sometimes have different roles and participate with their skills in this process. The technical terms with their different meanings in different areas of expertise can lead to different interpretation. Hence the need for a lexicon application which makes it possible to share the same understanding of the term in the field [15].

Lexicons may appear in different forms, such as glossary or dictionary data in a very simple form or LEL in a more or less elaborate form. Then, the lexicon is not a simple requirement for the quality of a process, but it is also a source of reference for stakeholders. In [5,11,15] found that the lexicon (LEL) like ontology represents only the overall view of a system, and does not allow stakeholders to represent the detailed conceptual level of the system. In addition, properties or features of every concept are combined in the notion or in the behavioral response of the LEL symbol. Furthermore, the LEL does not provide any information regarding the format and size of each data properties. This is why many research projects have proposed to divert or transform the LEL to describe or reveal details about the domain concepts or issues. These researches always define the context of the system in a UofD [9], which should be considered as the first phase of the process of the application domain model construction [4], and the models generated during this phase are used as the input of the next phase. Some researchers use the UofD to divert the LEL [9] and the LEL will be used in turn to obtain the scenario [1, 9], the ontology [7], the UML class diagram [12, 13] and the use case [2, 8]. To derive or transform the LEL into other conceptual model, an analysis followed by conversion will be conducted and the LEL becomes the initial stage of this process [14].We see that the necessary information for derivative models are already in the UofD which is the origin of the LEL. This results in the need for a more elaborate LEL which shows the conceptual level and the characteristics of each concept.

In this paper, we therefore propose a specific strategy to build the requirement specification in the form of lexicon rich in information characterized by one input with four outputs, called eLEL (section 2). The eLEL makes it possible to display the concepts of an application in the form of both natural and conceptual language oriented model. Then in section 3, we propose a construction process based on a series of heuristics in order to find the symbols, their structures and their meaning, and the various steps are undertaken to obtain the complete requirement model. Before concluding and saying a few words about various perspectives (Section 6), we will present a case study.

## 2. eLEL OR A MORE ELABORATED LANGUAGE EXTENDED LEXICON

A more elaborate Lexicon Extended Language or eLEL is a set of symbols (signs) according to the [3] definition. Indeed according to [3], a sign is something which can be interpreted to substitute something. According to [6], a sign is defined as an entity that has an expression and a content, but an eLEL symbol is a simple coding system with five entities: terms, notions, behavioral responses, attributes and methods. In eLEL, the terms classified in four categories like LEL [9] are divided into four types: the object, the subject, the verb and the state. Each eLEL term can be described according to its type by the heuristics presented in Tables 1, 2, 3 and 4. Table 1 describes the subject type eLEL.

Table 1 : Subject type eLEL

| eLEL symbol type | Subject |
|---|---|
| Description | This is an active entity with relevant roles in the application. The subject can be a person, a software component or another system with which interactions will occur. |
| Notion | Describes: Who is the subject? What are its characteristics? What are the objects it manipulates? |
| Behavioral responses | Describes: what is the definition of the functions performed by the subject? |
| Attribute | It shows a characteristic of the subject, such as the code, the technical wording or name etc. It is defined by its name, its code, its size, its nature or its type and description. Thus, a subject might have different attributes. |
| Method | Represents an operation which makes it possible to manipulate an attribute |

Table 2 shows the heuristics associated with the symbol of the eLEL object type.

Table 2: Object Type eLEL

| eLEL symbol type | Object |
|---|---|
| Description | This is a passive entity manipulated by a subject type eLEL. |
| Notion | Describes: what is the purpose? What are their characteristics? What are the other objects with which it is related? |
| Behavioral response | Describes: What are the actions applied to this object. ? |
| Attribute | It shows a characteristic of an object, such as the code, the technical wording or name etc. It is defined by its name, its code, its size, its nature or its type and description. Thus an object may have different attributes. |
| Method | Represents the action used to access an object or modify it. |

Table 3 shows the heuristics associated with the object of the verb type eLEL

Table 3: Verb type eLEL.

| eLEL symbol type | Verb |
|---|---|
| Description | Describes a feature that is performed by the subjects with its impacts on the operational environment. |
| Notion | Describes: who intervenes when an event happens or takes place? What is the object manipulated by the subject? What is the purpose or the objective to be achieved? |
| Behavioral response | Describes: what is the environmental impact, the resulting state and the satisfactory conditions for achieving the objective or purpose? |
| Attribute | Represents the subjects or objects affected by the verb. |
| Method | What are the actions to be taken by the subject on the objects participating in the realization of the objective or goal to be achieved? |

Table 4 shows the heuristic of state type eLEL object.

International Journal of Computer Science, Engineering and Applications (IJCSEA) Vol.5, No.6, December 2015

Table 4: State type eLEL.

| eLEL object type | State |
|---|---|
| Description | It is characterized by considerable attributes that contain values at different times during the running of the system. |
| Notion | Describes: what it represents? What actions led to it? |
| Behavioral response | Describes: how to identify other states that can be reached by the current state? |
| Attribute | Represents the subjects or objects that change the state. |
| Method | Represents actions taken to produce this state. |

## 3. eLEL BUILDING PROCESS

Building eLEL requires the following two principles [7]: the first one is to maximize the lexicon term used to describe the notion, the behavioral response, the method and the attribute of a new term, this is the principle of closure or circularity. The second is to minimize the use of terms outside the UofD. If it is unavoidable, it is necessary to ensure that the vocabularies used belong to the basic vocabulary of the natural language and as far as possible with a clear mathematical representation, this principle is called minimal vocabulary principle. In [7] mentioned that the eLEL term elucidation like the LEL is always done through a combination of technical elucidations. Then, during this technical elucidation, heuristics are used to find the relevant words in the UofD as well as the terms used in a very specific goal. In [7] also proposed that first, the initial terms are listed, then the reading and analysis of detailed documents can be used to assign the notion, the behavioral response, the attribute and the method of the recorded words. The eLEL construction process consists of thirteen steps based on the process proposed by [7, 11, 13]:

• Step 1: This is the identification of the main source of information in the UofD.

• Step 2: This is the identification of relevant terms in the UofD using a set of technical elucidation such as the statistical method that integrates the occurrence of a word or a noun phrase. Each identified term that seems to have a special meaning is listed.

• Step 3: This is the classification of each term by typology. Each term must be classified as an object, subject, verb or state (see Tables 1, 2, 3, 4).

• Step 4: This is the description of the notion and the behavioral response of the term by type:

- For the subject, describe the notion of the term by answering the questions, "Who is the subject" What are their characteristics? And what are the objects it manipulates? "And describe the behavioral response of the term by answering the question: " what is the definition of the functions performed by the subject? "(See Table 1).
- For the object, describe the notion of the term by answering the questions: "what is the object?" What are its characteristics? What are the related objects? "And describe the behavioral response of the term by answering the question" What are the actions applied to this object? "(See Table 2).

- For the verb, describe the notion of the term by answering the questions: "Who intervenes?" What is the object manipulated by the subject? And what is the purpose or goal to be achieved? "And describe the behavioral response by responding to the questions:" What is the environmental impact, the resulting state and the satisfactory conditions to achieve the goal? "(See Table 3).

- For the state, describe the notion of the term by answering the questions: "What does it represent? And what are the actions that lead to it? "And describe the behavioral response of the term by answering the question: "how to identify another state that can be achieved in addition to the current status? "(See Table 4).

In describing the notion and behavioral response of the terms in the lexicon, we should follow the principles of closure and minimal vocabulary.

• Step 5: For the subject, extract the attributes from the responses to the question: "what are their characteristics? " (See Table 1). After other terms have been referenced by the closure principle, this step identifies each name and defines them as attributes or properties of the term.

• Step 6: For the object, extract the attributes from the responses to the question: "what are their characteristics? " (See Table 2). What is not referenced as another term by the closure principle, this step identifies each name and defines them as attributes or properties of this term. The methods to access or modify each attribute are defined by adding respectively in each attribute name the prefixes GET and SET.

• Step 7: Each term attribute obtained in the fifth and sixth steps must be analyzed, its code is given, then its definition, format and size are deduced. The code, the definition, the format and the size characterize each attribute.

• Step 8: The actions or the methods of the term must be deduced from each entry of the behavioral response of the term classified as subject.

• Step 9: This step is about finding the method parameters obtained in step 8. Each term classified as a verb comes from the behavioral response classified as subject, therefore it describes all the data required to complete the behavior. The rules that model the actors and each term resources classified as verb parameter methods obtained in step 8 originated from the entry of the behavioral response origin of this symbol. The parameters characterize the eLEL symbol of verb types, so they are its attributes and the verb itself is the method or action. We must then analyze each term attribute obtained, its code is given, then we deduce its definition, format and size. The code, the definition, the format and the size obtained characterize each attribute.

• Step 10: This step is about the term classified as state. The verb type term is associated with the state of the environment before and after its execution, it defines the condition that must be previous to the implementation and to the situation which must be accomplished after the execution. The attributes and methods of a state type term are defined as follows: The method is that the eLEL verb type which starts the event, the attributes include the parameters used by the eLEL verb to trigger the event. Then we must analyze each term attribute obtained, its code is given, then its definition, format and size are deduced. The code, definition, the format and the size obtained characterize each attribute.

• Step 11: This step involves the verification of the lexicon that uses the inspection strategy invented by [9].
• Step 12: This step involves the validation [10] of the lexicon executed by the actors of the UofD using a writing technique.

• Step 13: This last step ensures that the eLEL symbols made valid in step 12 is linked in pairs through the *principle of circularity* :

-Each created eLEL symbol is assigned to a concept called *created element.* The concept *created element* is characterized by two attributes, its name and an eLEL symbol and composed of another concept known as *number of created elements* which is characterized by the minimum and maximum occurrence of associated eLEL symbol.

- After that, the entries of each eLEL symbol (*source symbol*) must be analyzed in the order. Then, we must detect another eLEL symbol in relation (*target symbol*).

-The relationship between symbols is represented by a concept called *circularity*. The *circularity* concept is characterized by its *name*, a *source symbol*, a *target symbol* and a *created element* concept which contains a couple of *names* of the two *created element* concepts corresponding to the *source symbol* and the *target symbol***.**

## 4. PROPOSITION OF EXTRACTION RULE

The proposed eLEL model is very rich in information. And it is possible to convert or extract it from other models. We present here as an example some eLEL transformation rules in class diagram.

Rule 1: All eLEL symbol classified as subject and object corresponds each to a UML class.

Rule 2: We extract the attributes of each eLEL symbol obtained in Rule 1 to constitute the attributes of each corresponding class. The code, format and size of each symbol attribute eLEL concerned respectively describe the name, format and size of the attribute of the corresponding class.

Rule 3: We extract the methods of each eLEL symbol obtained in Rule1 to constitute the methods of each corresponding UML class.

Rule 4 : Extraction *relationship* between eLEL object in *association* between class. This transformation is applicable for both subjects and objects. The *circularity* concept obtained through Step 13 of the eLEL construction process is extracted to form the *association* between each UML class.

Rule 5 : The concept *number of created elements* obtained through step 13 of the eLEL construction process is extracted to form the *cardinalities* of the classes obtained through rules 1, 2, 3 and 4.

## 5. CASE STUDY

In this section we instantiate the eLEL construction process, presented in section 3, with the process of issuing a birth certificate in the civil registry management system and the issuance of a birth certificate. Then, we transform the resulting eLEL models obtained in a class diagram.

The process of issuing a birth certificate begins with the declaration of birth made by the declarant by completing a systematic birth declaration form and then the actual issuance of the certificate by the civil status officer. The issuance of a birth certificate is a service offered by the registry office which can deliver the birth certificate of the newborn declared by the declarant. We present below some examples of eLEL symbols that belong to this case study. A description is provided for each symbol. A process of issuing a birth certificate is a classical administrative application in the field of civil status. In this case, the eLEL is used to represent each symbol in each category of the typology.

In Madagascar, the declaration of birth must realized within 12 days after the date of birth. The civil status officer requests a confirmation of the new birth from the declarant, it is after this confirmation that the issuance is done. If the 12-day period is exceeded, the birth certificate will not be issued or the process has to be repeated.

International Journal of Computer Science, Engineering and Applications (IJCSEA) Vol.5, No.6, December 2015## 5.1. Construction of eLEL

Step 1: We get the UofD as the main source of information. From the "vital events declaration form" and the "newborn identification process, the declaration and registration systemization of vital events".

Example 1 (excerpt from the UofD):

*"**1**-A birth declaration form is made up of the Region, the Municipality, the District, the Neighborhood, the information about the newborn, the information about the parents and the information about the declarant. **2**- A declaration sheet contains the place of birth, the date of birth, the name and the signature of the declarant and the civil status officer. **3**-The declarant, the civil status officer, the newborn and the parents of the newborn are person beings. **4**- Each person may have a name, a first name, a date of birth etc. **5**-The declarant fills in the vital events form. **6**-The civil status officer receives the vital events form. **7**-The civil status officer makes the birth certificate. **8**-The civil status officer issues the birth certificate. **9**-A birth certificate is issued. "*

Step 2: We obtained a list of candidate terms for the construction of eLEL objects:
Example 2 (from the list of candidate terms)

1. Municipality.
2. Declarant
3. District.
4. Birth declaration form.
5. The civil status officer issues the birth certificate.
6. The civil status officer prepares the birth certificate.
7. Newborn.
8. Birth certificate issuance process.
9. Neighborhood.
10. Receives the vital events form.
11. Region.
12. Fill in the vital events form.
13. Civil status officer.
14. A birth certificate is issued.

Step 3: We obtained the term classification by eLEL typology.

Table 5: List of terms by classification.

| Type | Candidate terms |
| --- | --- |
| Subject | Declarant, birth certificate issuance process, civil status officer. |
| Object | Municipality, District, Birth declaration form, Newborn neighborhood, Region |
| Verb | To issue the birth certificate, to prepare the birth certificate, to receive the civil status form, to fill in the civil status form |
| State | The birth certificate issued |

Step 4: We got the description of the notion and the behavioral response of each term classified by eLEL type (see Tables 6, 7, 8, 9 and 10).

Table 6: Description of an eLEL symbol type subject.

| eLEL Symbol | Birth certificate issuance process |
|---|---|
| Type | Subject |
| Notion | This is an information system for **a birth certificate issuance process.** <br> It is made up of the birth declaration form. <br> It is made up of the **acknowledgement of receipt   civil status officer.** <br> It is made up of the **birth confirmation**. <br> It is made up of an application **calculating the number of days between** the date of **the declaration of birth** and date of birth. <br> It is made up of a **request for confirmation** for the **declarant**. <br> It contains the number of the certificate. <br> It contains the date of registration in **the civil register** <br> It contains the surname and the first name of the **civil status officer.** <br> It contains the surname and name of the **declarant**. |
| Behavioral response | It can make it possible to **declare the birth**. <br> It can make it possible **to receive the birth declaration**. <br> It can make it possible to **calculate the   days** between the date of the **birth declaration** and the date of birth. <br> It can make it possible to **request a confirmation from the declarant.** <br> It can make it possible to **confirm the declaration of birth** <br> It can make it possible to **register the declaration of birth** in the **civil registry**. <br> It can make it possible to **prepare the birth certificate**. <br> It can make it possible to **issue the birth certificate**. |

Table 7: Description of the concept and behavioral response of a subject term.

| eLEL Symbol | Declarant |
|---|---|
| Type | Subject |
| Notion | This is a person who **declares the birth**. <br> It is an entity characterized by a name, a first name, an address, and the quality of the **declarant.** <br> It provides the **Region of birth**, **the district of birth, the municipality of birth**, **the information about the newborn**, **the information about the newborn's father**, **the information about the newborn's mother**, the issuance date of the certificate, the date of birth, the place where the certificate was made. |
| Behavioral response | It fills in the information about **the birth Region**, **the District of birth**, the **municipality of birth, about the neighborhood where he was born, about the newborn, the newborn's father and about the newborn's father.** |

Table 8: Description of the concept and behavioral response of a term like verb.

| eLEL Symbol | To issue a birth certificate |
|---|---|
| Type | Verb |
| Notion | The **civil status officer** must **issue the birth certificate** following the **process of issuing the birth certificate.** |
| Behavioral response | The **birth certificate is delivered** to the **declarant** on a date fixed by the **civil status officer** |

Table 9: Description of the concept and behavioral response of a term like object.

| eLEL Symbol | Birth certificate declaration form |
|---|---|
| Type | Object |
| Notion | This is **a form completed by the declarant to declare a birth**. It contains **the birth region**, the birth district, the neighborhood where he was born, **information about the newborn**, **information about the father**, **information about the mother** and **information about the declarant**. It contains the number of the certificate, the month, the year, the hour, the minute, the day (in the morning or in the evening), the birth date, the birth place, the declaration date and the **declarant** and the civil status officer's signature. |
| Behavioral response | It makes it possible to **declare the birth**, to **receive the birth declaration**, to **calculate the number of days**, to **request confirmation**, to **confirm the declaration**, to **register the declaration**, to **prepare the birth certificate** and to **issue the birth certificate** |

Table 10: Description of the concept and behavioral response of a state like term.

| eLEL Symbol | Birth certificate issued |
|---|---|
| Type | State |
| Notion | The situation in which the **birth certificate** of **the newborn** is delivered at the end of the **process for issuing a birth certificate.** It is conducted by the action **deliver the birth certificate.** |
| Behavioral response | The **Date of the issuance of the first birth certificate is fixed. The birth certificate is issued.** |

Step 5: We got the attributes of subject like terms (Tables 11 and 12).

Table 11: Representation of the Attributes of subject like terms.

| eLEL : Birth certificate issuance process |
|---|
| Type : Subject |
| **Attributes** |
| Date of the declaration, The Civil Status Officer's name, The Civil Status Officer's name, The Declarant's name, The Declarant's first name. |

Table 12: Representation of the attributes of subject like terms.

| eLEL : Declarant |
|---|
| Type : Subject |
| **Attributes** |
| The declarant's name, the declarant's first name, the address, the declarant's first name, the declarant's quality, the date of the birth certificate, date of birth, place of birth |

Step 6 and 7: Applying step 6, we got the attributes and methods of the words classified as objects and step 7 provides the features of each attribute, such as its code, its definition, its format and size. The terms classified object are all obtained in this step.

Table 13: eLEL symbol of Object type.

| eLEL Symbol | Birth certificate declaration form |
|---|---|
| Type | Object |
| Notion | This is a **form completed by the declarant** to **declare a birth**. It consists of the **birth region**, the **birth District**, the **neighborhood where he was born**, the **information about the newborn**, the **information about the father**, the **information about the mother** and the **information about the declarant**.<br>It contains the number of the certificate, the month, the year, the hour, the minute, the day (in the morning or in the evening), the date of birth, the place of birth, the date of the statement and the **declarant** and the **civil status officer**'s signatures |
| Behavioral response | It can make it possible **to declare the birth**, to **receive the declaration of birth**, to **calculate the number of days**, to **request a confirmation**, to **confirm the statement**, to i**nclude the declaration**, to **make the document** and **to issue the birth certificate.** |

| **Attributes** | | | | |
|---|---|---|---|---|
| Name | Code | Definition | Format | Size |
| Number of the certificate | num_cert_birth | Number of the birth certificate | Digit | 6 |
| Month | birth_month | Month of birth | Digit | 2 |
| Year | birth_year | Year of birth | Digit | 4 |
| Hour | birth_hour | Hour of birth | Digit | 2 |
| Minute | birth_min | Minute of birth | Digit | 2 |
| Day | birth_day | Birth date of the newborn | Digital | 2 |
| Birth Date | birth_date | Birth Date of the newborn | Date | 8 |

| **Methods** |
|---|
| getBirthCertNum() |
| setBirthCertNum() |
| getBirthMonth() |
| setBirthMonth() |
| getBirthYear() |
| setBirthYear() |
| getBirthHour() |
| setBirthHour() |
| getBirthMin() |
| setBirthMin() |
| getBirthDay() |
| setBirthDay() |
| getBirthDate() |
| setBirthDate() |

Step 8: We obtained the methods of each term classified subject (see table 14 and 15).

Table 14: Methods of a subject term.

| eLEL Symbol | Birth certificate issuance process |
|---|---|
| Type | Subject |
| **Methods** | |
| DeclareBirth() | |
| ReceiveDeclaration() | |
| CalculateNumberDays() | |
| RequestConfirmation() | |
| ConfirmDeclaration() | |
| Registereclaration() | |
| MakeBirthCertificate() | |
| DeliverBirthCertificate() | |

Table 15: Methods of a subject term.

| eLEL Symbol | Declarant |
|---|---|
| Type | Subject |
| **Methods** | |
| EnterRegion() | |
| EnterDistrict() | |
| EnterMunicipality() | |
| EnterNeighborhood() | |
| EnterNewbornInfo() | |
| EnterFatherInfo() | |
| EnterMotherInfo() | |
| EnterDeclarantInfo() | |

Step 9: We got the methods of terms classified as subject as well as the attributes and methods of the terms classified as verbs. The terms of subject and verb types are defined in this step.

Table 16: Description of an eLEL symbol type subject.

| eLEL Symbol | Birth certificate issuance process |
|---|---|
| Type | Subject |
| Notion | This is an information system to perform **a birth certificate issuance process.** |
| | It contains the **birth declaration form.** |
| | It contains the **acknowledgement of receipt** by **the civil status officer.** |
| | Il contains the **birth confirmation**. |
| | It contains an application **calculating the number of_days between** the date of **the declaration of birth** and date of birth. |
| | It consists of a **request for confirmation** for the **declarant**. |
| | It contains the number of the certificate. |
| | It contains the date of registration in **the civil register** |
| | It contains the surname and the first name of **civil status officer** |
| | It contains the surname and name of the **declarant**. |
| Behavioral response | It can enable us to **declare the birth**. |

International Journal of Computer Science, Engineering and Applications (IJCSEA) Vol.5, No.6, December 2015

|  | It can enable us **to receive the birth declaration**. It can enable us to **calculate the number of days** between the date of the **birth declaration** and the date of birth. It can enable us to **request a confirmation from the declarant** It can enable us to **confirm the declaration of birth** It can enable us to **register the declaration of birth** in the **civil registry**. It can enable us to **prepare the birth certificate**. It can enable us to **issue the birth certificate**. |  |  |  |
|---|---|---|---|---|
| **Attributes** | | | | |
| Name | Code | Definition | Format | Size |
| Birth declaration Date | declaration_date | Date of the birth declaration | Date | 8 |
| Name of the civil status officer | civ_stat_officer_name | Name of the civil status officer | Text | 25 |
| First name of the civil status o | civ_stat_firstname | First name of the civil status officer | Text | 25 |
| Name of the declarant | declarant_name | Name of the declarant | Text | 25 |
| First name of the declarant | declarant_firstname | First name of the declarant | Text | 25 |
| **Methods** | | | | |
| DeclareBirth() | | | | |
| ReceiveDeclaration() | | | | |
| CalculateNumberDays() | | | | |
| RequestConfirmation() | | | | |
| ConfirmDeclaration() | | | | |
| RegisterDeclaration() | | | | |
| EstablishBirthCertificate() | | | | |
| DeliverBirthCertificate() | | | | |

Table 17: Description of an eLEL symbol type subject.

| eLEL Symbol | Declarant |
|---|---|
| Type | Subject |
| Notion | This is a **person** who **declares the birth**. It is an entity characterized by a name, a first name, an address, and the quality of the **declarant.** It provides the **Region of birth, the district of birth, the municipality of birth, the information about the newborn, the information about the father of the newborn, the information about the newborn's mother,** the issuance date of the certificate, the date of birth, the place where the certificate was made. |
| BehavioralResponse | It fills in the information about **the birth Region, the District of** |

| | | birth, the municipality of birth, about the neighborhood where he was born, about the newborn, the newborn's father and about the newborn's father. | | |
|---|---|---|---|---|
| **Attributes** | | | | |
| Name | Code | Definition | Format | Size |
| Name | Name | Name of declarant | Text | 25 |
| First Name | Firstname | First name of declarant | Text | 25 |
| Adress | Adress | Adress of declarant | Text | 65 |
| Quality | quality | Quality of declarant | Text | 15 |
| Birth certificate Date | birth_cert_date | Date of the birth certificate | Date | 8 |
| Birth date | birth_date | Date of birth | Date | 8 |
| Place of the Certificate | cert_place | Place of the certificate | Text | 25 |
| **Methods** | | | | |
| EnterRegion() | | | | |
| EnterDistrict() | | | | |
| EnteMunicipality() | | | | |
| EnterNeighborhood() | | | | |
| EnterNewbornInfo() | | | | |
| EnterFatherInfo() | | | | |
| EnterMotherInfo() | | | | |
| EnterDeclarantInfo() | | | | |
| | | | | |

Table 18: Description of an eLEL object of a verb type.

| eLEL Symbol | Issue the birth certificate | | | |
|---|---|---|---|---|
| Type | Verb | | | |
| Notion | The **civil status officer** must **issue the birth certificate** following the **process of issuing birth certificates.** | | | |
| Behavioral response | **The birth certificate is delivered** to the **declarant** on a date fixed by the **civil status officer** | | | |
| **Attributes** | | | | |
| Name | Code | Definition | Format | Size |
| The birth certificate | copy_birth_cert | Copy of the birth certificate | Complex | 1 |
| Declarant | declarant | Declarantof the birth | Complex | 1 |
| Civil status officer | civ_stat_officer | Civil status officer | Complex | 1 |
| **Methods** | | | | |
| DeliverBirthCertificate(). | | | | |

Step 10: We have defined the terms of state type

Table 19: Description of an eLEL symbol type state.

| eLEL symbol | Copy of the birth certificate issued |
|---|---|
| Type | State |
| Notion | The situation in which the **birth certificate** of **the newborn** is delivered at the end of the **process for issuing the birth certificate.** <br> It is conducted by the action **deliver the birth certificate.** |
| Behavioral response | **The date of the issuance of the first birth certificate is fixed. The birth certificate is issued.** |
| **Attributes** | | | | |

| Name | Code | Definition | Format | Size |
|---|---|---|---|---|
| Birth declaration form | declaration_form | Birth certificate declaration | Complex | 1 |
| Declarant | declarant | Declarant of the birth | Complex | 1 |
| Civil status officer | civ_stat_officer | Civil status officer | Complex | 1 |

| **Methods** |
|---|
| DeliverBirthCertificate() |

Step 11 and 12: To complete the construction process, an expert study has been carried out by linguists to verify the description of the eLEL (step 11). Finally, all the parties concerned have also done the validation of the eLEL symbols (step 12).

Step 13: By applying Step 13, we have the relationship between symbols made valid as well as its occurrencies.

## 5.2. Transformation of the eLEL obtained in a class diagram

Rule 1: We got the list of UML classes corresponding to each eLEL symbol of object and subject type.

Table 20: List of obtained UML class candidates.

| Type | eLEL Object | UML class |
|---|---|---|
| Subject | Declarant, Process of issuing birth certificate, Civil Status Officer. | Declarant, Process of issuing birth certificate, Civil Status Officer. |
| Object | Municipality, District, Birth certificate declaration form, Newborn, Region, Newborn's mother, Newborn's father. | Municipality, District, Birth certificate declaration form, Newborn, Region, Newborn's mother, Newborn's father. |

Rule 2 and 3: By applying rule 2 and 3 respectively, we got the attributes and methods for each UML class from rule 1.

International Journal of Computer Science, Engineering and Applications (IJCSEA) Vol.5, No.6, December 2015

Table 21: Example of class UML, methods and attributes.

| UML classes | Attributes | Methods |
| --- | --- | --- |
| Declarant | name, firstname, adress, quality, date_birth_certif, date_birth | EnterRegion(),EnterDistrcit(),EnterMunicipality(),EnterNeighborhood(),EnterNewborn(),EnterFather(),EnterMother(),Enterdeclarant(). |
| Process of issuing the birth certificate | date_declaration, name_civ_stat_officer, firstname_civ_stat_officer | Declare_birth(),receive_declaration(),calculateNumberDays,requestConfirmation(), registerDeclaration(), establishBirthCertificate(), issuebirthcertificate() |
| Birth certificate declaration form | birth_certif_numb, birth_month, birth_year etc. | getNum, setNum etc… |

Rule 4 et 5: We have the UML class diagram abstract model after the extraction (Figure 3) corresponding to the eLEL symbol subject and object (Figure 2) by translating rules 1-5 in ATL transformation (Atlas Transform Language). (Figure 1) show the ATL transformation of rule 5 eLEL extraction into UML class diagram.

```
83 --------R5 CreatedElement2Cardinality---------
84 rule ElementcreatedNumber2Cardinality{
85     from r:MMeLEL!CreatedElement,
86          m:MMeLEL!NomberOfCreatedElement
87     (r.numberofcreatedelement=m)
88     to  pr:MMUML!Property
89     (
90         name<-r.name.replaceAll(' ','_'),
91         class<-MMeLEL!Circularity.allInstances()->select(x|x.createdelement->collect(y|y.createdsymbol)->includes(r.createdsymbol))
92         ->collect( y |y.createdelement).flatten()
93         ->select(y|y.createdsymbol<>r.createdsymbol).first().createdsymbol,
94         lower<-m.lower,
95         upper<-m.upper
96
97     )
98     }
99 -----------End R5---------------------
```

Figure 1. ATL transformations of the *number of created element* concept of rule 5 eLEL extraction into UML classes *cardinality*.

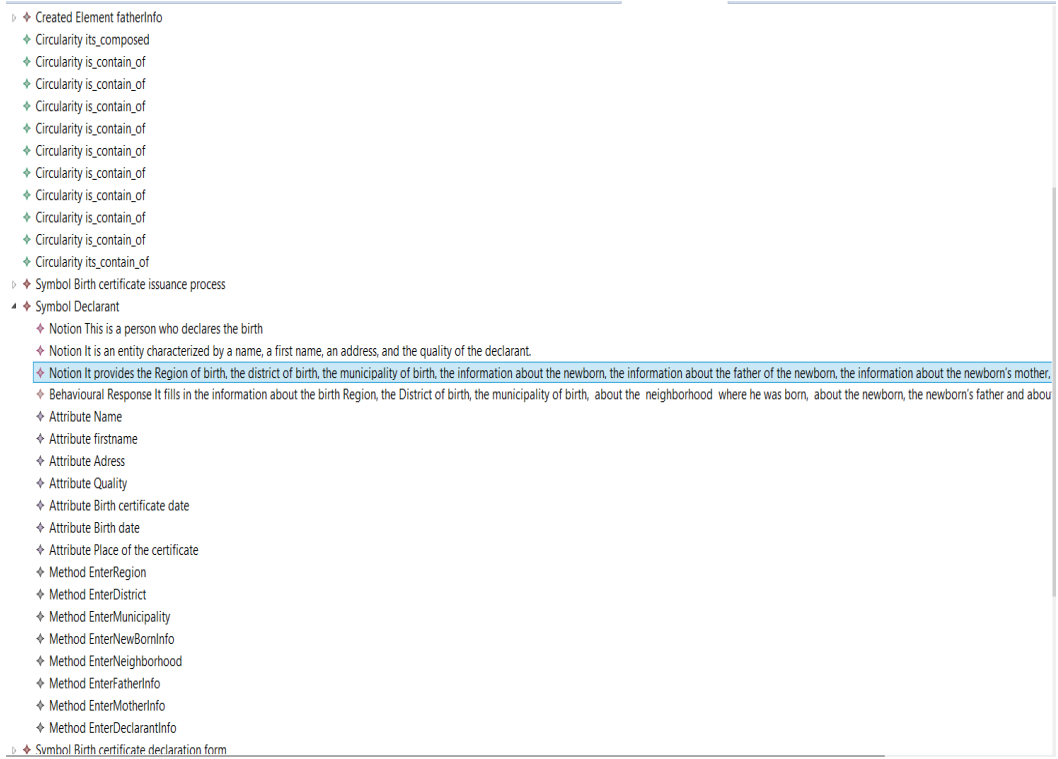

Figure 2. Ecore Model Sample Reflective for the eLEL symbol model.

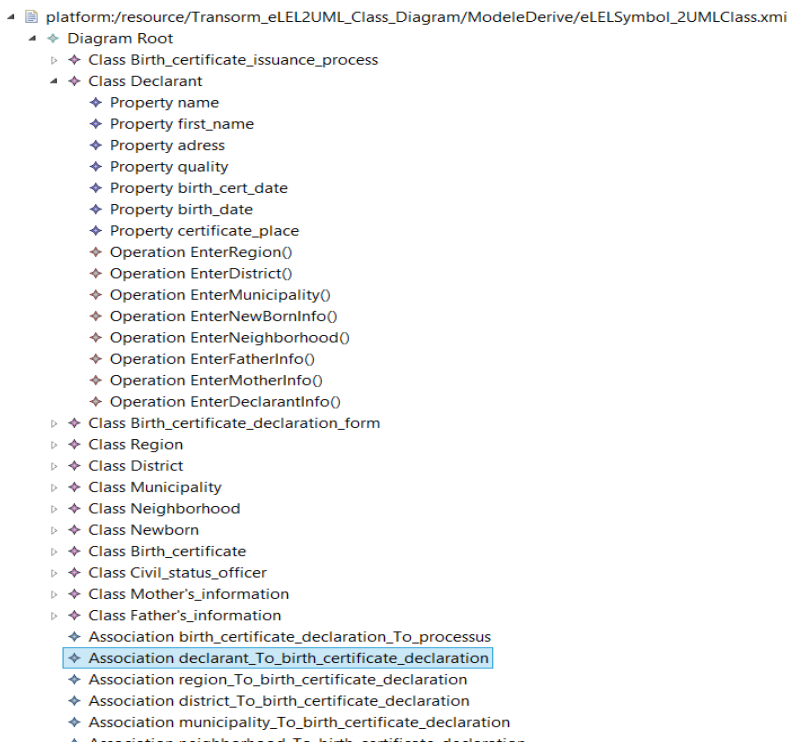

Figure 3. Ecore Model Sample Reflective for UML class diagram abstract.

## 6. CONCLUSION AND FUTURE WORK

In this paper, we have proposed a conceptual model enriched in natural language which enables us to describe in more details the concept in the field of eLEL. Then we have proposed a modeling process using the eLEL model. The eLEL consists of an entity, attributes, methods, notions and behavioral responses. The entity is the term or the concept itself, the attributes and methods are the conceptual view, the notions and the behavioral responses define the semantics of natural language of each model. The heuristics and the stages of the process of building eLEL used in this article allow us to build an eLEL, given the case study we conducted. The eLEL construction process steps are so important in requirements engineering that they must be developed carefully. The eLEL construction steps process provide a systematic and logical way to define the conceptual view of an application model in natural language. The approach provided in this article enables us to describe the different concepts of the UofD by explicitly defining structural and behavioral aspects. The eLEL model is a conceptual model of a natural language application domain. So it is a model closer to the raw original model because it comes directly from the UofD. Thanks to its wealth of information, it is possible to easily extract it from application requirements as well as data dictionary of a domain.

And as perspectives of this work, we identified the following:

-The use of eLEL in MDM (Master Data Management), to design, implement an information system, and to map the core business of an enterprise.
-The extraction of an eLEL symbol from different UML diagrams such as the use case diagram, the activity diagram, the state diagram etc.
-The derivation of scenarios.
-The computerization of the birth certificate.
-Website to create.

## ACKNOWLEDGEMENTS

I would like to thank Pr. Josvah Paul RAZAFIMANDIMBY and Dr. Thomas MAHATODY for their most support and encouragement. They kindly read my paper and offered invaluable detailed advices on grammar, organization, and the theme of the paper.

## AUTHORS

Jean Luc RAZAFINDRAMINTSA  is a student PhD  at the  Research Institute School for Computer Modelisation and the Laboratory of Mathematical and Computer Applied to the Development Systems at the University of Fianarantsoa Madagascar. The main research topics are modelling Master Data Management,  modelling Service Oriented Architecture, modelling Enterprise Architect and Business Intelligence by using elaborate natural language model oriented requirements.

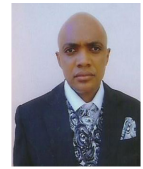

Thomas MAHATODY is a PhD at the National School for Computer Engineering  and the Director of TIC at the University of Fianarantsoa Madagascar.

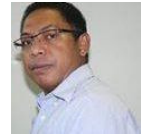

Josvah Paul RAZAFIMANDIMBY is a full Professor at the University of Madagascar and the Director of the Laboratory of Mathematical and Computer Applied to the Development Systems at the University of Fianarantsoa Madagascar.

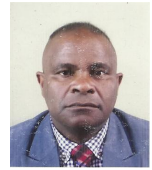